\documentstyle[prd,aps]{revtex}
\begin{document}
\title{\bf The Block Spin Renormalization Group Approach and\\
           Two-Dimensional Quantum Gravity }
\author{\bf Ray L. Renken }
\address{Department of Physics, University of Central Florida,
Orlando, Florida, 32816 }
\date{April 25, 1994}
\maketitle
\mediumtext
\widetext
\begin{abstract}
\begin{quotation}

A block spin renormalization group approach is proposed for the dynamical
triangulation formulation of two-dimensional quantum gravity.  The idea is to
update link flips on the block lattice in response to link flips on the
original lattice.  Just as the connectivity of the original lattice is
meant to be a lattice representation of the metric, the block links are
determined in such a way that the connectivity of the block lattice represents
a block metric.  As an illustration, this approach is applied to the Ising
model coupled to two-dimensional quantum gravity.  The correct critical
coupling is reproduced, but the critical exponent is obscured by unusually
large finite size effects.

\end{quotation}
\vskip .25in
\end{abstract}
\section{ Dynamical Triangulations and Quantum Gravity }

Dynamical triangulations are by now well established as a discretization
of two-dimensional quantum gravity, at least when the central charge
of the matter it is coupled to is less than one \cite{review}.  A
two-dimensional triangulation is characterized by its points $i$, its links
$<ij>$, and its triangles, $<ijk>$.  This information is uniquely determined
by the adjacency matrix, $G_{ij}$, which defines the nearest-neighbor pairs:
\begin{equation}
G_{ij} = \left\{  \begin{array}{ll}
                    1 & \mbox{ i,j are nearest neighbor} \\
                    0 & \mbox{ otherwise. }
                    \end{array}
         \right.
\end{equation}
All links are defined to be of length $a$, the lattice spacing, which is
usually taken to be one.  Since the triangles are therefore all equilateral,
the coordination number at a site $i$, $q_i$, is related to the curvature,
$R_i$
\begin{equation}
R_i = \pi (6 - q_i)/q_i.
\end{equation}
Six is the average coordination number on a regular triangulation which
has zero curvature.  The adjacency matrix acts like a metric in the sense
that it tells which points are near and which are far away.  For points that
are not nearest neighbors, the geodesic distance is defined to be the length
of the shortest path between those points.  In a quantum theory of gravity,
a sum over all possible metrics is required to compute the partition function.
On the lattice, this becomes a sum over all possible triangulations.  If an
additional statistical mechanical variable is placed on the nodes to represent
matter, then a sum over all possible matter configurations is required for each
triangulation.  In a number of cases, these quantum gravity plus matter systems
can be solved analytically.

The Ising model illustrates this and is the only example that will be
discussed in this paper.  In the absence of quantum gravity, i.e. on a fixed
regular triangulation, the partition function is defined as
\begin{equation}
Z = \sum_{S} \exp\left( k\sum_{<ij>}S_iS_j \right)
\end{equation}
where the sum is over all configurations of the spins, $S_i = \pm 1$.  This,
the usual Ising model \cite{Baxter}, has a second order phase transition at
$k = k^c$ where
\begin{equation}
k^c = \frac{1}{2} \sinh^{-1} {1 \over \sqrt 3} = 0.2746 \cdots \label{kwithout}
\end{equation}
and it has critical exponent
\begin{equation}
\nu = 1 \label{nuwithout}
\end{equation}
where $\nu$ is defined by the behavior of the correlation length, $\xi$, near
the critical coupling
\begin{equation}
\xi (k - k^c) \approx (k - k^c)^{-\nu}. \label{nudef}
\end{equation}
In the presence of quantum gravity, the definition of the partition function
includes the sum over triangulations (labeled by $T$)
\begin{equation}
Z = \sum_T \sum_{S} \exp\left( k\sum_{i,j}G_{ij}(T)S_iS_j \right).
\end{equation}
This theory also has a second order phase transition
\cite{ising1,Burda}, this time at
\begin{equation}
k^c = -\frac{1}{2} \ln \tanh \frac{1}{2} \ln \left( {108\over 23} \right)
= 0.2162\cdots
\label{kwith}
\end{equation}
with critical exponent
\begin{equation}
\nu d = 3. \label{nuwith}
\end{equation}
The exponent now appears in combination with the Hausdorff dimension, $d$,
which is dynamically determined.  This combination appears because the
correlation length (the object whose behavior $\nu$ describes) is a distance
whereas in quantum gravity direct control over length scales is lost.  The
relation between volume (still under control) and distance is given by the
Hausdorff dimension.
It has been shown that for matter with central charge less than one, the
scaling dimensions at a critical point of the pure matter theory are
``dressed" by quantum gravity so that the critical behavior of the two
theories is related \cite{KPZ}.

Not all models of interest can be solved exactly.  Numerical methods provide
another option.  The key fact that makes simulations of dynamical
triangulations practical is that any triangulation (for a given topology) can
be reached from any other triangulation by a sequence of link flips
\cite{BKKM}.  A link flip is defined in figure 1 and is the replacement of a
link with another connecting the two other points in the two triangles it
shares.  Matter can be simulated in the usual way.

In numerical simulations of lattice theories for high energy physics, the
primary difficulty is often dealing with critical phenomena.  When the lattice
spacing is taken to zero, the correlation length (in physical units) is zero
unless the correlation length in lattice units is infinite.  Thus the location
of continuous phase transitions and the study of their universal behavior
are generically problems of interest for such theories.  Finite size scaling
provides one method of computing critical couplings and exponents.  This
method has been successfully applied to systems coupled to quantum gravity
such as the crumpling transition \cite{crump} and the Ising model
\cite{ising3}.

Another method, the block spin renormalization group approach, is the subject
of this paper.
In this method, an effective theory is constructed by partitioning
the lattice into blocks and averaging the degrees of freedom within each block.
Starting with a critical theory and iterating this procedure produces a
sequence of theories which flow toward a fixed point.  Critical exponents can
be calculated using expectation values of operators determined at this fixed
point.  This approach was first developed on spin theories (hence the name)
and applied also to lattice gauge theories.
It has not yet been applied to (the more recently invented)
dynamically triangulated random surfaces.  The objective of this paper is to
apply the block spin approach to this newer type of theory.

\section{ The Block Spin Renormalization Group Approach }
First, consider in more detail the block spin approach
(see \cite{rg1,rg2,rg3,rg4,rg5,rg6,rg7}).
Denote a lattice spin variable by $S$ so that the partition function is
\begin{equation}
Z = \sum_{S} e^{H[S]}.
\end{equation}
Now average over blocks of spins as illustrated in figure 2.  A square lattice
is used, but the same scheme works on a triangulation which can be obtained
by drawing in all of the up and to the right diagonals.  Label each block
by $m$ and assign it a new spin $t_m$.  More generally, a probability can be
assigned for every possible value of $t_m$:
\begin{equation}
P(t_m) = K(t_m,|S|_m)
\end{equation}
where $|S|_m$ denotes the set of spins within the block.  Given a configuration
of initial spins, the probability for a configuration of block spins is
\begin{equation}
P[t] = \prod_m K(t_m,|S|_m).
\end{equation}
Multiplying by the probability of the initial configuration and summing over
all possible ititial configurations gives the total probability of obtaining
the block configuration, which can be viewed as resulting from an effective
theory
\begin{equation}
e^{H^\prime[t]} = \sum_{S} \prod_m K(t_m,|S|_m) e^{H[S]}.
\end{equation}
The effective theory has (for the scenario of figure 2) twice the original
lattice spacing.  If the physics is unchanged by this operation, the
correlation length is, in lattice units, half its previous value.
The physics is unchanged if the partition functions of the blocked and
original theories are identical.  This is assured if
\begin{equation}
\sum_{t_m} K(t_m,|S|_m) = 1. \label{unitarity}
\end{equation}
An example of a solution to this equation is
\begin{equation}
K(t_m,|S|_m) = \delta (t_m - f(|S|_m)).
\label{kernel}
\end{equation}
For instance, the ``majority rule" for the Ising model is of this form: for
an odd number of spins if the majority are up the block spin is up.
Otherwise it is down.  For an even number of Ising spins, a tie-breaker is
required (trivially modifying the previous equation).

In general, if the block spins are the same type of variable as the original
spins and if the blocked lattice is isomorphic to the original, then $K$ can
be applied iteratively to produce a sequence of effective Hamiltonians.
Distances must be rescaled at each step.  If the scale factor is $b$, the
ratio of the number of points in the original lattice, $N$, to the number
of points in the block lattice, $N^\prime$, is
\begin{equation}
N/N^\prime = b^D
\end{equation}
where $D$ is the dimensionality of the lattice.  The correlation length is
reduced by $b$:
\begin{equation}
\xi^\prime = \xi / b. \label{rescale}
\end{equation}
More abstractly, the renormalization group transformation acts on the space
of Hamiltonians
\begin{equation}
H^\prime = R[H].
\end{equation}
If this transformation has a fixed point
\begin{equation}
R[H^*] = H^*
\end{equation}
then
\begin{equation}
\xi^* = \xi^* / b
\end{equation}
which has solutions $\xi^* = 0$ and $\xi^* = \infty$.  The former corresponds
to a (trivial) system at infinite temperature.  The latter corresponds to a
(nontrivial) critical point.  There are many possible renormalization group
transformations for a given system and not all of them iterate to a nontrivial
fixed point for the critical Hamiltonian at hand.  Aside from the constraint of
Eq. (\ref{unitarity}) and the requirement that the transformation reduce the
number of degrees of freedom, it is also necessary to ask that it be apt, i.e.
that it focuses on the critical phenomena of interest.

Now perturb $H^*$ with an operator $O$ so that the system is slightly off of
criticality.  It is convenient to reparametrize the scale factor
\begin{equation}
b = e^l
\end{equation}
in order to discuss differential changes of scale (now mediated by a
transformation $R_l$).  Assume that
\begin{equation}
{d\over dl}(H^* + O) \equiv \lim_{l\rightarrow 0} {R_l[H^*+O^*] - (H^*+O^*)
\over l} = LO + \cdots
\end{equation}
where $L$ is a linear operator and nonlinear terms (denoted by the dots) are
neglected.  If the eigenoperators of $L$ are defined by the equation
\begin{equation}
L O_i^* = \lambda_i O_i^*
\end{equation}
and the $O_i^*$ form a complete set of operators in the neighborhood of $H^*$,
then for $H_0$ in that neighborhood
\begin{equation}
H_0 = H^* + \sum_i h_i O_i^*.
\end{equation}
The evolution of the coefficients (called linear scaling fields) is determined
by
\begin{equation}
{dh_i(l)\over dl} = \lambda_i h_i(l)
\end{equation}
implying
\begin{equation}
h_i(l) = h_ie^{\lambda_i l} \label{*1}
\end{equation}
so that
\begin{equation}
H_l = H^* + \sum_i h_i e^{\lambda_i l} O_i^*
\end{equation}
where $H_l$ represents the effective Hamiltonian after the scale is changed
by $b = e^l$.  The $\lambda_i$ are critical exponents: if $\lambda_i > 0$ the
$i^{\rm th}$ perturbation grows ($O_i^*$ is relevant) while if $\lambda_i < 0$
the $i^{\rm th}$ perturbation diminishes ($O_i^*$ is irrelevant).  If
$\lambda_i = 0$, $O_i^*$ is marginal.

To make contact with traditional scaling ideas, consider a system with only
one relevant operator whose corresponding linear scaling field is the coupling,
$k$.  Take all of the irrelevant couplings to be zero.  Then
Eq. (\ref{rescale}) can be written as
\begin{equation}
\xi(k - k^c) = e^l \xi(e^{\lambda l} (k - k^c)).
\end{equation}
If $l$ is chosen so that
\begin{equation}
e^l \propto (k - k^c)^{-1/\lambda}
\end{equation}
then
\begin{equation}
\xi(k - k^c) = {\rm (const)} (k - k^c)^{-1/\lambda}
\end{equation}
which is the same as Eq. (\ref{nudef}) after making the identification
\begin{equation}
\nu = 1 / \lambda.
\end{equation}

How can this approach be realized in a numerical setting?  In such a setting,
it is expectation values of functions of products of the original spins
\begin{equation}
<O> = \sum_{S} O(S) e^{H[S]}
\end{equation}
that are readily accessible.  Expectation values in the blocked theory can
be computed with the original spins using the definition of blocked spins
\begin{eqnarray}
<O(t)> & = & \sum_{t} O(t) e^{H^\prime [t]} \\
       & = & \sum_{t} O(t) \sum_{S} \prod_m K(t_m,|S|_m) e^{H[S]}.
\end{eqnarray}
If $K$ is a delta-function or majority rule then each configuration of spins
specifies a configuration of block spins so that the sum on $t$ is
determined.

On an infinite lattice, the criticality of an initial Hamiltonian could be
verified by observing that expectation values of arbitrary operators approach
fixed values as the renormalization group procedure is iterated.  In practice,
a computationally manageable lattice is typically small, especially after
several iterations of the transformation.  Finite size effects spoil the
matching.  This problem can be circumvented by comparing expectation values
obtained
on a lattice of $N$ points blocked $m$ times to those obtained on a lattice
of $N^\prime$ points blocked $m-1$ times with $N/N^\prime = b^D$.  The two
block lattices are the same size so that differences in expectation values
are due to the difference in the Hamiltonians alone.

Once the couplings are near their critical values, they can be systematically
improved.  Near the fixed point Hamiltonian
\begin{equation}
<O_i> = <O_i>_{H = H^*} +
\sum_j{\partial <O_i>_{H = H^*} \over \partial k_j}\delta k_j
+ \cdots
\end{equation}
where the $k_j$ are the relevant bare parameters in the theory and
$\delta k_j = k_j - k_j^c$.  Formulating, as
described above, the difference between expectation values taken from original
lattices of different sizes blocked down to lattices the same size gives
\begin{equation}
<O_i^{(n)}>_L - <O_i^{(n-1)}>_S =
\sum_j \left[
{\partial <O_i^{(n)}>_L \over \partial k_j} -
{\partial <O_i^{(n-1)}>_S \over \partial k_j}
\right]
\delta k_j \label{deltak}
\end{equation}
where $L$ (large) denotes an initial lattice with some volume $N$ and $S$
(small) denotes an initial lattice with volume $N/b^d$.  The
superscripts indicate the blocking level.  This equation predicts the
adjustment to the bare parameters necessary to achieve criticality.  The term
in brackets can be computed from expectation values using expressions of the
form
\begin{equation}
 {\partial <O_i^{(n)}> \over \partial k_j^{(m)}} =
<O_i^{(n)}O_j^{(m)}> - <O_i^{(n)}><O_j^{(m)}>. \label{*2}
\end{equation}
The critical exponents can also be obtained numerically.  If, after $n$
iterations of the renormalization group transformation
\begin{equation}
H^{(n)} = H^* + O_i (k_i^{(n)} - k_i^*)
\end{equation}
then
\begin{equation}
R[H^* + (k_i^{(n)}-k_i^*)O_i] = H^* + T_{ij}^*(k_i^{(n)}-k_i^*)O_j + \cdots
\end{equation}
where now the scale factor, $b$, is fixed and the stability matrix, $T_{ij}^*$,
gives
\begin{equation}
(k_i^{(n)}-k_i^*) = T_{ij}^* (k_j^{(n-1)}-k_j^*)
\end{equation}
or
\begin{equation}
T_{ij}^* =
\left. {\partial k_i^{(n)}\over \partial k_j^{(n-1)}}\right|_{H^*}.
\label{stability}
\end{equation}
Diagonalizing this matrix gives linear scaling fields, $h$, (linear
combinations of the $k_i$) and eigenvalues $\Lambda$ which obey
\begin{equation}
h_i^{(n)} = \Lambda_i h_i^{(n-1)}.
\end{equation}
Comparing with Eq.\ (\ref{*1}) reveals that these eigenvalues are related
to the critical exponents,
\begin{equation}
\Lambda_i = b^{\lambda_i}. \label{critexp}
\end{equation}
The stability matrix can be determined from correlations of the type in
Eq.\ (\ref{*2}) by using the chain rule
\begin{equation}
 {\partial <O_i^{(n)}> \over \partial k_j^{(n-1)}} =
\sum_l {\partial k_l^{(n)} \over \partial k_j^{(n-1)}}
{\partial <O_i^{(n)}> \over \partial k_l^{(n)}}.
\label{chain}
\end{equation}
Arbitrary couplings are implicated, so $T_{ij}^*$ is an infinite matrix.  In
practice, only a finite number of expectation values can be handled, so
it must be truncated.  This is a source of error.

\section{ A Block Spin Transformation for Dynamical Triangulations }
The block spin renormalization group approach has been successfully applied
to spin models and gauge theories.  How can it be applied to dynamical
triangulations?  Since the Hausdorff dimension is dynamically determined,
there is no direct access to the length scale.  Further, the degrees of
freedom do not consist of elements of an algebra or manifold residing on a
regular lattice so that they can simply be averaged in some way and projected
back onto the algebra or manifold.  A block spin approach for triangulations
will have to look a little different.

Imagine a triangulation and another triangulation that is somehow a result
of blocking the original one.  The blocked lattice should have fewer
points, but it should still be a triangulation.  If the original triangulation
is regular (implying toroidal topology), it is clear how to arrange this.  For
example, see figure 3.  The volume (number of points) is decreased by a factor
of four.  Now, any other triangulation (of the torus) can be reached from the
original triangulation by a sequence of link flips.  Likewise, any other block
triangulation can be reached from the blocked one in figure 3 by a sequence of
block link flips.  The central idea of this paper is this:  any rule that
dictates when a block link should be flipped in terms of when the original
links are flipped is equivalent to a definition of a renormalization group
kernel, $K$.

In order for a kernel to be apt it must preserve the important physics.
Physically, the triangulation is viewed as a discretization of a euclidean
spacetime with the adjacency matrix acting as a metric: points connected by
a link are ``close", while points not connected by a link are not close.
The block lattice is produced by marking a subset of points on the original
lattice and then connecting them (defining a blocked adjacency matrix) to
form the blocked lattice.  If the blocked adjacency matrix is to act as an
effective metric on the block lattice, it should indicate which block sites
are close and which are further away.
A link on the block lattice has two neighboring triangles just as in
figure 1.  Define the geodesic distance between block points $x$ and $y$,
$r_{xy}$, as the minimum number of links that must be traversed on the
underlying lattice to get from one point to the other.  If block link flips
are made whenever
\begin{equation}
r_{ad} < r_{bc} \label{geodesic}
\end{equation}
then the block lattice preserves the idea of a metric.  This now defines a
block spin renormalization group transformation for dynamical triangulations.
It will turn out to be apt.

In order to handle matter coupled to this
system, a block spin algorithm must be specified for the matter as well.
Only Ising spins will be discussed in what follows.  A rule must be specified
defining which nodes are associated together in a block.  Presumably one and
only one of the nodes that are specified as block nodes will appear in each
block.  Then a rule (such as the majority rule) must be given relating the
block spin to those spins in the block.  For a regular triangulation (like that
of figure 3), the block node itself and the three neighboring nodes in the
forward directions (right and up) could be defined as being in the block.
This is the scheme of figure 2.
On a dynamical lattice such an assignment is not always possible.
There are often points on the original lattice that are greater than distance
one from any block node.  This means that either nodes further than distance
one from the block node must be included in the block, or less than four nodes
must sometimes be used.

The definition of block spin used here is as follows.  Each block node is in
its own block.  Each block node is then allowed to pick (randomly) one
nearest-neighbor that has not yet been picked (two different block nodes
sometimes
share a neighbor).  This is repeated twice so that three neighbors are chosen
if there is no contention.  If, at some point in the selection process,  there
are no neighbors that have not been spoken for, no node is selected and that
block has fewer than four spins.  The majority rule then determines the block
spin.  The weakness in this procedure is that the coordination number of a
site influences the selection of the spins in a block and therefore also the
determination of the block spin.  This could effectively contribute a relevant
perturbation to the Hamiltonian, spoiling the matching of expectation values.
Expectation values in the gravitational sector should not be influenced,
because the selection of block links in no way depends on the selection of
block spins.  In the calculations discussed below, there is a small effect
present in the spin sector (at the highest blocking level considered) but not
in the gravitational sector that might by explained in this way.

These ideas are implemented using the Ising model coupled to quantum gravity
with the critical value of the coupling constant as given in Eq. (\ref{kwith}).
The triangulations are updated using the link flip algorithm and the
spins are updated using the Wolff algorithm \cite{wolff}.  The block links
are updated using the algorithm based on Eq. (\ref{geodesic}).  This is
computationally intensive because it involves the calculation of the distance
between block nodes in terms of the links of the underlying lattice.  This
distance can be large at the lower blocking levels.  Instead of calculating
both of the distances involved in Eq. (\ref{geodesic}), it is quicker to just
determine which is smaller.  This can be done in a loop that labels all of
the neighbors of the four block sites involved, then all of their neighbors
and so on until the neighbors from one of the diagonal pairs meets.  The block
link should join this pair.  Five to ten passes through all of the links of
the block lattice are made with this procedure in order to implement the
triangulation blocking algorithm.  The block spin procedure is then iterated
by treating the block lattice as an original lattice.

All original lattice sizes are chosen so that a regular triangulation would
block down to a $3\times 3$ torus.  This choice is made to ease comparisons
between
the cases with and without quantum gravity.  The minimal triangulation of a
torus requires seven points, so nine points is not wastefully large.  The
renormalization group transformation defined in this paper rescales the volume
by a factor of four, so possible volumes for the initial lattices are nine
times powers of four.  The volumes used are $144$, $576$, and $2304$.  These
lattices can be renormalized twice, three times, and four times, respectively.
Runs on these lattices involved $10^5$ passes through each lattice where
a pass is defined as $16$ sweeps through each link of the lattice with the
link flip algorithm along with either $200$ Wolff updates (for the smaller
two lattices) or $800$ (for the largest).

Performing the renormalization group procedure at the critical coupling
should produce an expectation value of any given operator that approaches a
constant up to finite size effects which are controlled by only comparing
numbers obtained on effective lattices of the same size.  Three operators
are used from the spin sector and three from the gravity sector.  $O_1$ is
defined as the product of spins at the opposite ends of a link.  $O_2$ is
defined as the product of spins at the opposite ends of the conjugate link,
the link that would result from a flip.  $O_3$ is defined as the product of all
four spins involved in $O_1$ and $O_2$.  $O_4$, $O_5$, and $O_6$ are defined
just like the previous three, except instead of using the spin, the
coordination number minus six is used.  This gives information about the
gravitational sector.  Another gravity sector operator computed ($O_7$) is the
maximum
coordination number on the lattice.  Table I shows the resulting expectation
value of the Ising term ($O_1$) as a function of the volume and blocking level,
$n$.
This table demonstrates both the finite size effects and the way they are
countered by always comparing numbers obtained on lattices of the same size.
Reading across the rows, the numbers vary due to finite
size effects.  Reading along diagonals, down and to the left, the numbers are
all obtained from effective lattices that are the same size and, in the bottom
two diagonals,  they (nearly) approach a constant.
The bottom number in each column is from a lattice with volume nine.  If
Eq. (\ref{deltak}) is used to predict how far the input coupling is from the
true critical one, the prediction for $\delta k = k - k^c$ is
\begin{equation}
\delta k = 0.0002(3)
\end{equation}
comparing the $V=144$ and $V=576$ data at the highest blocking levels (two
and three, respectively).
This number should be zero, since the input coupling was the known critical
coupling.
Results from similar runs, but with the system slightly off of criticality,
are shown along with this data in figure 4.  For large enough iterations
of the renormalization group transformation and for input couplings close
enough to the critical value, the results from Eq. (\ref{deltak}) would fall
on a straight line (as indicated in the figure).  The plotted points, obtained
with a relatively small number of iterations, are close to this line.

Thus, at this level of iteration, the critical coupling can be estimated
quite accurately.  What happens at higher levels of iteration?  Here, a hint
of trouble begins to appear.  At $k = k^c$, the prediction for $\delta k$ is
\begin{equation}
\delta k = 0.0008(3) \label{bad}
\end{equation}
comparing the $V=576$ and $V=2304$ data at the highest blocking levels.  This
number is statistically different from zero and signals some kind of trouble.
This same difficulty can be seen in the data in table I, where the $n = 4$
number does not match as well with the $n = 3$ number on the same diagonal
as the $n = 3$ and $n = 2$ numbers did.  This result
is not improved by longer runs or more effort updating the block links.
It could be a result of correlations induced in the block spins by the
blocking algorithm as discussed earlier.  Evidence that the problem is in the
spin sector rather than the gravity sector is given by table II, which gives
the behavior of all of the operators blocked to a lattice with volume nine.
All of the gravity sector operators match within statistics.
So, while the blocking procedure for triangulations appears to be satisfactory,
Eq. (\ref{bad}) signals a possible flaw in the blocking procedure for matter
that needs to be investigated more carefully.

With this caveat regarding the $n = 4$ spin operators, what about critical
exponents?  Table III lists estimates for $1/\nu d$ obtained using Eqns.
(\ref{stability}) and (\ref{chain}).  Since the volume is
being rescaled here and not a length scale, Eq. (\ref{critexp}) must be
rewritten slightly.  If the volume rescaling factor is $v = b^d$, then
\begin{equation}
\Lambda_i = v^{\lambda_i / d}
\end{equation}
so that $1 / \nu d$ is obtained from the maximum eigenvalue of the stability
matrix using
\begin{equation}
{1\over \nu d} = {\ln \Lambda_{\rm max} \over \ln v}
\end{equation}
with $v = 4$.
The matrix $T_{ij}^*$ can be truncated with different numbers of operators
and those estimates are listed in each column of table III.  The operator
order for truncations is $O_1$, $O_3$, $O_2$, and $O_4$, Satisfactory error
estimates could not be obtained when $O_5$ and $O_6$ were included.
Looking at the table, it is immediately clear that there
are effects from both the finite size of the lattice (as seen
by looking across rows) and the limited number of iterations of the
renormalization group transformation.  It is normal to have these effects,
but they are unusually large here.  Compare this data to that from the
three-dimensional Ising model as studied in \cite{Wilson}.  There, using
the regularity and the smallness of the finite size effects they are able
to give an infinite lattice estimate for each blocking level.  Then, giving
some thought to the magnitude of the subleading exponent, they are able to
extrapolate to a large number of blockings and obtain an estimate of
$1/\nu d$ in excellent
agreement with others.  In their analog of Table III, the difference between
the largest and smallest numbers in the table is no more than 30\%.  Here the
large variation precludes such an analysis. It is possible that with sufficient
data on larger lattices a similar procedure would produce the correct
result (Eq. (\ref{nuwith})) $1/\nu d = 1/3$.

There are systems where finite size effects and effects due to too few
blockings are so small that the detailed procedure of \cite{Wilson} is
virtually unnecessary.  The two-dimensional Ising model in the absence of
quantum gravity is an example.  By taking the program that produced the
previous results and deleting the line that calls the routine that updates
the triangulations, it is easy to reproduce the correct pure Ising exponents.
For
instance, starting with a $24\times 24$ lattice, setting the coupling to the
value given by Eq. (\ref{kwithout}), and using two operators in
the truncation of the stability matrix, the sequence of estimates of
$1/\nu d$ for $n = 1, 2, {\rm and~} 3$ is, respectively, $0.467(7)$, $0.51(2)$,
and $0.498(7)$.  These compare favorably with the exact answer (from
Eq. (\ref{nuwithout})), $0.5$.

It should be possible to do block spin calculations similar to the one
described in this paper, but on arbitrary topologies and with more general
volume rescalings.
Simply mark points on the original lattice and create a
triangulation with them in order to provide an initial block lattice with
the desired volume and topology.  One way to do this, starting with an initial
lattice of volume $V = N$ and genus $g$, is to knock out (do the inverse of
barycentric subdivision on) $n$ three-fold coordinated nodes and their
associated
triangles to produce a volume $V^{\prime} = N - n$.  Once this initial block
lattice exists, the simulation can be carried on as above by updating the block
lattice according to the geodesic rule, Eq. (\ref{geodesic}).
This defines the renormalized lattice.
The matter field is another issue and is likely to be problematic if
$V^{\prime}/V \approx 1$.  For $V^{\prime}/V = 1/p$ with $p$ an integer, a
majority rule could be used to average the site and $p-1$ randomly chosen
neighbors.

\acknowledgements
I thank Simon Catterall, Elbio Dagotto, John Kogut, and Ken Wilson for
helpful discussions.

\vskip .5in
\centerline{List of Figures}
\vskip .5in
\begin{enumerate}
\item A link flip.
\item A simple blocking scheme for spins.  The nodes of the original lattice
are represented by the intersections of the straight lines.  Block spins are
formed by averaging (as with Eq. (\ref{kernel})) the spins grouped with the
rounded squares.
\item A blocking scheme for a regular triangulation.
\item Predicted versus actual values for $\delta k = k - k^c$ obtained using
Eq. (\ref{deltak}). A line with slope one is drawn for comparison.
\end{enumerate}
\vskip .5in
\centerline{List of Tables}
\vskip .5in
\narrowtext
\begin{table}
\begin{tabular} {cddd}
n & $V = 2304$ & $V = 576$  &  $V = 144$  \\ \hline
0 & 0.4946(1)  & 0.4989(2)  &  0.5093(4)  \\
1 & 0.2304(4)  & 0.3066(5)  &  0.3971(7)  \\
2 & 0.2450(7)  & 0.3517(9)  &  0.481(1)   \\
3 & 0.350(1)   & 0.482(2)   &             \\
4 & 0.489(2)   &            &             \\
\end{tabular}
\caption{ The expectation value of the Ising term as a function of the number
of iterations of the renormalization group transformation, $n$, and the volume
of the initial lattice. }
\end{table}
\mediumtext
\begin{table}
\begin{tabular} {cddd}
operator & $V = 2304, n = 4$ & $V = 576, n = 3$  &  $V = 144, n = 2$  \\ \hline
$O_1$ &  0.489(2) &  0.482(2)  &  0.481(1)  \\
$O_2$ &  0.488(2) &  0.482(2)  &  0.479(1)  \\
$O_3$ &  0.306(2) &  0.293(2)  &  0.289(1)  \\
$O_4$ & -0.365(4) & -0.367(2)  & -0.368(1)  \\
$O_5$ &  0.347(8) &  0.355(5)  &  0.360(2)  \\
$O_6$ &  0.20(4)  &  0.15(1)   &  0.038(8)  \\
$O_7$ &  7.879(5) &  7.880(2)  &  7.867(1)  \\
\end{tabular}
\caption{ Expectation values of a variety of operators (defined in the text)
as a function of the
initial lattice's volume for $n$ such that the blocked lattices have volume
nine in each case. }
\end{table}
\narrowtext
\begin{table}
\begin{tabular} {cddd}
  n   & $V = 2304$& $V = 576$ &  $V = 144$ \\ \hline
  1   & -0.269(3) & -0.068(3) &  0.083(2)  \\
      & -0.260(4) & -0.080(4) &  0.076(2)  \\
      &           &           &            \\
  2   &  0.417(2) &  0.378(3) &  0.342(3)  \\
      &  0.426(2) &  0.403(3) &  0.385(5)  \\
      &  0.455[1] &  0.432[3] &  0.410[5]  \\
      &  0.455[1] &  0.432[3] &  0.410[5]  \\
      &           &           &            \\
  3   &  0.493(1) &  0.407(3) &            \\
      &  0.534(2) &  0.463(6) &            \\
      &  0.545(1) &  0.476(6) &            \\
      &  0.545(1) &  0.476[5] &            \\
      &           &           &            \\
  4   &  0.449(3) &           &            \\
      &  0.505(5) &           &            \\
      &  0.513(6) &           &            \\
      &  0.513[7] &           &            \\
\end{tabular}
\caption{ Estimates for $1/\nu d$ as a function of the volume of the initial
lattice, the level $n$ (from Eq. (41)), and the number of operators
used in the truncation
of the stability matrix.  From one to four operators are used and the results
are listed down for each blocking level. Errors listed in parentheses were
obtained by binning the data into at least twenty bins.  Those errors listed
in square brackets were obtained using fewer bins. }
\end{table}

\begin{references}

\bibitem{review} V. A. Kazakov and A. A. Migdal, Nucl. Phys. B311, 171 (1988).

\bibitem{Baxter} {\it Exactly Solved Models in Statistical Mechanics},
                 R. J. Baxter, Academic Press, 1984.

\bibitem{ising1} D. V. Boulatov and V. A. Kazakov, Phys. Lett. B186,
                 379 (1987).

\bibitem{Burda} Z. Burda and J. Jurkiewicz, Acta. Phys. Polon. B20,
                9949 (1989).

\bibitem{KPZ} V. G. Knizhnik, A. M. Polyakov, and A. B. Zamolodchikov,
              Mod. Phys. Lett. A3, 819 (1988).

\bibitem{BKKM} D. V. Boulatov, V. A. Kazakov, I. K. Kostov, and A. A. Migdal,
               Nucl. Phys. B275, 641 (1986).

\bibitem{crump} R. L. Renken and J. B. Kogut, Nucl. Phys. B354, 328 (1991).

\bibitem{ising3} S. M. Catterall, J. B. Kogut, and R. L. Renken, Phys. Rev.
                 D45, 2957 (1992).

\bibitem{rg1} K. G. Wilson, Rev. Mod. Phys. 47, 773 (1975).

\bibitem{rg2} K. G. Wilson, in {\it Recent Developments in Gauge Theories},
              ed. G 't Hooft (Plenum Press, N. Y. 1980).

\bibitem{rg3} R. H. Swendsen, Phys. Rev. Lett. 47, 1775 (1981).

\bibitem{rg4} R. H. Swendsen, J. Appl. Phys. 53, 1920 (1982).

\bibitem{rg5} C. Rebbi and R. H. Swendsen, Phys. Rev. B21, 4094 (1980).

\bibitem{rg6} M. A. Novotny, D. P. Landau, and R. H. Swendsen,
              Phys. Rev. B26, 330 (1982).

\bibitem{rg7} M. E. Fisher, {\it Scaling, Universality and Renormalization
              Group Theory}, in {\it Critical Phenomena} Lecture Notes in
              Physics, Vol. 186, ed F. J. W. Hahne (Springer-Verlag, Berlin,
              1983).

\bibitem{wolff} Ulli Wolff, Phys. Rev. Lett. 62, 361 (1989).

\bibitem{Wilson} G. S. Pawley, R. H. Swendsen, D. J. Wallace, and K. G. Wilson,
                 Phys. Rev. B29, 4030 (1984).

\end{references}
\end{document}